\documentclass[11pt]{article}
\usepackage{amsfonts,amsmath,amsthm,amssymb,
graphicx,mathrsfs}
\usepackage[top=3cm,bottom=3cm,left=4cm,right=3cm]{geometry}

\usepackage{booktabs}

\usepackage[width=.9\textwidth, labelfont=bf,
            textfont=it]{caption}

\usepackage{algorithm,algorithmicx,algpseudocode}
\algblockdefx{MRepeat}{EndRepeat}{\textbf{Repeat}}{}
\algnotext{EndRepeat}

\usepackage{hyperref}
\hypersetup{  colorlinks=true,
    linkcolor=blue,
    filecolor=magenta,      
    urlcolor=cyan,
    citecolor=blue}

\def\hat{\widehat}
\def\tilde{\widetilde}

\theoremstyle{plain}
\newtheorem{theorem}{Theorem}
\newtheorem{lemma}[theorem]{Lemma}

\newtheorem{remark}{Remark}

\makeatletter
\renewcommand{\@biblabel}[1]{}
\renewenvironment{thebibliography}[1]
     {\section*{\refname}%
      \@mkboth{\MakeUppercase\refname}{\MakeUppercase\refname}%
      \list{}%
           {\labelwidth=0pt
            \labelsep=0pt
            \leftmargin1em
            \itemindent=-1em
            \advance\leftmargin\labelsep
            \@openbib@code
            }%
      \sloppy
      \clubpenalty4000
      \@clubpenalty \clubpenalty
      \widowpenalty1000%
      \sfcode`\.\@m}
\makeatother
\usepackage{breakcites}

\allowdisplaybreaks

\begin{document}

\title{Efficient Bayesian reduced rank regression using Langevin Monte Carlo approach}

\author{The Tien Mai$ ^{(1)} $}

\date{
\small
$^{(1)}$ Oslo Centre for Biostatistics and Epidemiology, Department of Biostatistics, 
\\
University of Oslo, Norway.
\\
Email: t.t.mai@medisin.uio.no
}

\maketitle

\begin{abstract}
The problem of Bayesian reduced rank regression is considered in this paper. We propose, for the first time, to use Langevin Monte Carlo method in this problem. A spectral scaled Student prior distrbution is used to exploit the underlying low-rank structure of the coefficient matrix. We show that our algorithms are significantly faster than the Gibbs sampler in high-dimensional setting. Simulation results show that our proposed algorithms for Bayesian reduced rank regression are comparable to the state-of-the-art method where the rank is chosen by cross validation.
\end{abstract}

\section{Introduction}

Reduced rank regression \cite{anderson1951estimating,izenman1975reduced,velu2013multivariate} is a widely used model in linear multivariate regression. In this model, the low-rank constraint is imposed on the coefficient matrix to promote estimation and prediction accuracy of multivariate regression. This low-rank structure is building upon the belief that the response variables are related to the predictors through only a few latent direction. This assumption is also useful to extend the model to high dimension settings \cite{bunea2011optimal}.

Variouse methods have been conducted for reduced rank regression based on Bayesian approach. A first study can be found in \cite{geweke1996bayesian} in bayesian econometrics which based on a low-rank matrix factorization approach. Since then, various works have been studied, for example,  \cite{kleibergen2002priors,corander2004bayesian,babacan2012sparse,schmidli2019bayesian,alquier2013bayesian} and developed to low-rank matrix estimation and completion \cite{lim2007variational,salakhutdinov2008bayesian,mai2015}. Moreover, Bayesian reduced rank regression has been successfully applied in other fields such as genomics \cite{marttinen2014assessing,zhu2014bayesian}. More recently, several works have expanded Bayesian methods for incoperating the sparsity into the reduced rank regression \cite{goh2017bayesian,chakraborty2020bayesian,yang2020fully}.

It is noted however that most works in Bayesian reduced rank regression (BRRR) usually employ conjugate priors as the conditional posterior distributions can be explitcitely derived that allow to implement Gibbs sampling \cite{geweke1996bayesian,salakhutdinov2008bayesian}. The details of these priors are reviewed and discussed in \cite{alquier2013bayesian}. Nevertheless, these Gibbs sampling approaches need to calculate a number of matrix inversions or singular value decompositions at each iteration and thus it will be costly and can slow down significantly the algorithm for large data. Different attempts, however based on optimization, have been made to address this issue including variational Bayesian methods and maximum a posteriori approach \cite{lim2007variational,marttinen2014assessing,yang2018fast}.

In this paper, we consider a different way for choosing the prior distribution on low-rank matrices rather than based on the traditional low-rank matrix factorization. More specifically, a scaled Student prior is used in our approach for which the rank of the coefficient matrix does not need to be prespecified as in the matrix factorization approach. This prior has been successfully used before in a context related to low-rank matrix completion \cite{yang2018fast} and image denoising \cite{dalalyan2020exponential}.

We develop, for the first time, a Langevin Monte Carlo (MC) approach in Bayesian reduced rank regression. The Langevin MC method was introduced in physics based on Langevin diffusions \cite{ermak1975computer} and it became popular in statistics and machine learning following the paper \cite{roberts1996exponential}. Recent advances in the study of Langevin Monte Carlo make it become more popular in practice \cite{dalalyan2017theoretical,durmus2017nonasymptotic,dalalyan2020sampling} and a promissing approach for Bayesian statistics \cite{durmus2019high}. 

More specifically, in this work, we first present a naive (unadjusted) Langevin MC algorithm and then a Metropolis–Hastings correction for Langevin MC is proposed. Interestingly, our implementations does not require to perform matrix inversion nor singular values decomposition and thus our algorithms can deal with large data set efficiently. Numerical results from these two algorithm are comparable to the frequentist approach for which the rank is chosen using cross validation. More particularly, we further show that our proposed Langevin MC algorithms are significantly faster than Gibbs sampler in high-dimensional settings, see Section \ref{sc_numerical}.

The paper is structured as follows. In Section \ref{sc_model} we present the reduced rank regeression model, then the prior distribution is defined together with some discussion regarding the low-rank factorization prior. In Section \ref{sc_LMC}, the implementations of the Langevin MC approach are given in details. Numerical simulations and a real data application are presented in Section \ref{sc_numerical}.  Some discussion and conclusion are given in Section \ref{sc_discuss} and \ref{sc_conclu} respectively.

\section{Bayesian reduced rank regression}
\label{sc_model}
\subsection{Model}
We observe two matrices $X$ and $Y$ with
\begin{equation}
Y =  X B + \mathcal{E}
\end{equation}
where $ Y\in \mathbb{R}^{n\times m} $ is the response matrix, $ X \in \mathbb{R}^{n\times p} $ is the predictor matrix, $B \in \mathbb{R}^{p\times m} $ is the unknown regression coefficient matrix and $\mathcal{E}$ is an $n\times m$ random noise matrix with $\mathbb{E}(\mathcal{E})=0$. 

We assume that the entries $\mathcal{E}_{i,j}$ of $\mathcal{E}$ are i.i.d. Gaussian $\mathcal{N}(0,\sigma^2)$ and $ \sigma^2 $ is given. In this case, note that the likelihood distribution is given by
 $$ 
 L(Y)  \propto \exp\left\{ - \frac{1}{2\sigma^2}\|Y-XB\|^2_F   \right\} 
 $$
 where $\|\cdot\|_F$ denotes the Frobenius norm, $\|M\|_F^2
= {\rm Tr}(M^T M)$.

The objective is to estimate the parameter matrix $B$. In many applications, it makes sense to assume that the matrix $B$ has low rank, i.e. ${\rm rank}(B)\ll \min(p,m)$. 

With a prior distribution $\pi(B) $, the posterior for the Bayesian reduced rank regression is
\begin{align*}
\mathcal{L}_n (B) \propto  L(Y)  \pi(B).
\end{align*}
Then, the Bayesian estimator is defined as
\begin{align}
\label{bayestimator}
\hat{B} = \int B \mathcal{L}_n (B).
\end{align}
Note that here we focus on estimating $B$ and thus $\sigma^2 $ is fixed. On can relax this assumption and place a prior distribution on $\sigma$.

\subsection{A low-rank promoting prior}
Borrowing motivation from a low-rank prior explored in recent works \cite{yang2018fast,dalalyan2020exponential}, we consider the following prior,
\begin{align}  
\label{prior_scaled_Student}
\pi(B) 
\propto 
\det (\lambda ^2 \mathbf{I}_{p} + BB^\intercal )^{-(p+m+2)/2}
\end{align}
where $ \lambda > 0 $ is a tuning parameter and $ \mathbf{I}_{p} $ is the $p\times p$ identity matrix . 

To illustrate that this prior has the potential to encourage the low-rankness of $B$, one can check that 
\begin{align*}
\pi(B) 
 \propto \prod_{j=1}^{m}  (\lambda ^2  + s_j(B)^2 )^{- (p+m+2)/2 },
\end{align*}
where $ s_j(B) $ denotes the $j$the largest singular value of $B$. It is well known that the log-sum function $ \sum_{j=1}^m \log (\lambda ^2  + s_j(B)^2) $ encourages a sparsity on $\{s_j(B)\}$, see \cite{candes2008enhancing,yang2018fast}. Thus the resulting matrix $B$ has a low-rank structrure, approximately.  

The following Lemma explains the reason why this prior is a spectral scaled Student prior distribution.
\begin{lemma}
\emph{\cite{dalalyan2020exponential} }
If a random matrix $B$ has the density distribution as in \eqref{prior_scaled_Student}, then the random vectors $B_i$ are all drawn from the $p$-variate scaled Student distribtuion $ (\lambda /\sqrt{3})t_{3,p} $. 
\end{lemma}

\subsubsection*{On low-rank factorization priors}
The first idea about a low-rank prior was carried out in \cite{geweke1996bayesian}. That is to express the matrix parameter $B$ as $B_{p\times m}=M_{p\times k}N_{m\times k}^\top$ with $k \leq \min(p,m) $. The prior is difined on $M$ and $N$ rather than on $B$ as
$$
\pi(M,N) \propto 
\exp\left\{ - \frac{\tau^2}{2} ( \|M\|^2_F + \|N\|^2_F)   \right\} 
$$
for some $\tau>0$. This prior allows to obtain explicit forms for the marginal posteriors that allows an implementation of the Gibbs algorithm to sample from the posterior, see \cite{geweke1996bayesian}. However, the downside of this approach is the problem of choosing $k$, the reduced rank, is not directly addressed. Thus, one has to perform model selection for any possible $k$, as done in \cite{ kleibergen2002priors,corander2004bayesian}.

Recent approaches focus on fixing a large $k$, e.g. $k = \min (p,m)$, then sparsity-promoting priors are placed on the columns of $M$ and $N$ such that most columns are almost null. So that the resulting matrix $B = MN^\top$ is approximately low-rank. This direction was first proposed in \cite{babacan2012sparse} in the context of matrix completion, but it still can be used in reduced rank regression. See \cite{alquier2013bayesian} for the details and dicussions on low-rank factorization priors.

With low-rank factorization priors, most authors simulate from the posterior by using the Gibbs sampler as the conditional posterior distributions can be explitcitely derived, e.g. in \cite{geweke1996bayesian,salakhutdinov2008bayesian}. However, these Gibbs sampling algorithms update the factor matrices in a row-by-row fashion and invole a number of matrix inverse operations at each iteration.  This is expensive and slow down the algorithm for large data set.

\section{Langevin Monte Carlo implementations}
\label{sc_LMC}

In this section, we propose to compute an approximation of the posterior with the scaled multivariate Student prior by a suitable version of the Langevin Monte Carlo algorithm. 
\subsection{Unadjusted Langevin Monte Carlo algorithm}
Let us recall that the log posterior is of the following form
$$
\log\mathcal{L}_n (B) 
= 
- \frac{1}{2\sigma^2} \|Y-XB  \|_F^2 
- \frac{p+m+2}{2} \log \det (\lambda^2 \mathbf{I}_{p} + BB^\intercal ),
$$
and consequently,
$$
\nabla  \log\mathcal{L}_n (B) 
= 
- \frac{1}{\sigma^2} X^\intercal (Y-XB)
- (p+m+2) (\lambda^2 \mathbf{I}_{p} + BB^\intercal )^{-1}B,
$$
We use the constant step-size unadjusted Langevin MC (denoted by LMC) \cite{durmus2019high}. It is defined by choosing an initial matrix $B_0$ and then by using the recursion
\begin{align}
\label{langevinMC}
B_{k+1} = B_{k} - h\nabla \log\mathcal{L}_n (B_k) +\sqrt{2h}\,W_k,\qquad
k=0,1,\ldots,
\end{align}
where $h>0$ is the step-size and $ W_0, W_1,\ldots$ are independent random matrices with i.i.d. standard Gaussian entries. The detail of the algorithm is given in the Algorithm \ref{lmc_algorithm}.

Note that a direct application of the Langevin MC algorithm \eqref{langevinMC} needs to calculate  a $p\times p$ matrix inversion at each iteration. This can slow down significantly the algorithm and might be expensive. However, one can easily verify that the matrix
$\mathbf{M} = (\lambda^2 \mathbf{I}_{p} + BB^\intercal )^{-1}B $ is the solution to the following convex optimization problem
$$
\min_{\mathbf{M} } \big\{\| \mathbf{I}_m- B^\top \mathbf{M}  \|_F^2 + \lambda^2\|\mathbf{M} \|_F^2\big\}.
$$
The solution of this optimization problem can be obtained by using the package 'glmnet' \cite{glmnet} (with family option 'mgaussian'). This does not require matrix inversion nor other costly operation. However, it is noted that in this case we are using the Langevin MC with approximate gradient evaluation, theoretical assessment of this method can be found in \cite{dalalyan2019user}.

\begin{algorithm}{}
\caption{LMC for BRRR}
\begin{algorithmic}[1]
\State \textbf{Input}: matrices $Y\in\mathbb{R}^{n\times m} ,X \in \mathbb{R}^{n\times p} $
\State \textbf{Parameters}: Positive real numbers $\lambda,h,T$ 
\State \textbf{Onput}: The matrix $\hat{B} $
\State \textbf{Initialize}: $B_0 \gets (X^\top X + 0.1\mathbf{I}_p)^{-1}X^\top Y;  \hat{B} = \mathbf{0}_{p\times m}$ 
\For{$k \gets 1$ to $T$} 
    \State Simulate $B_{k}$ from \eqref{langevinMC};
    \State  $\hat{B} \gets  \hat{B} +  B_{k}/T $
\EndFor
\end{algorithmic}
\label{lmc_algorithm}
\end{algorithm}

\begin{remark}
It seems that the Algorithm \ref{lmc_algorithm} looks like an iterative gradient descent for minimizing the penalized Gaussian log-likelihood with a penalty. However, Algorithm  \ref{lmc_algorithm}  computes the posterior mean and not a maximum a posteriori estimator. More precisely, our algorithm includes a final step of averaging.
\end{remark}

\begin{remark}
For small values of $h $, the ouput $\hat{B} $ is very close to the integral \eqref{bayestimator} of interest. However, for some $h$ that may not small enough, the Markov process is transient and thus the sum explodes \cite{roberts2002langevin}. To address this problem, one have to take a smaller h and restart the algorithm or a Metropolis–Hastings correction can be included in the algortihm. The Metropolis–Hastings approach ensures the convergence to the desired distribution, however, it greatly slows down the algorithm because of an additional acception/rejection step at each iteration. The approach by taking a smaller $h $ also slows down the algorithm but we keep some control on its time of execution.
\end{remark}

\begin{remark}
Based our observations from numerical studies in Section \ref{sc_numerical}, the initial matrix $B_0$ can also effect the convergence of the algorithm. We suggest using $ B_0 = (X^\top X + 0.1\mathbf{I}_p)^{-1}X^\top Y $ as a default alternative.
\end{remark}

\subsection{A Metropolis-adjusted Langevin algorithm}

Here, we consider a Metropolis-Hasting correction to the Algorithm \ref{lmc_algorithm}. This approach guarantees the convergence to the posterior. More precisely, we consider the update rule in \eqref{langevinMC} as a proposal for a new state,
\begin{align}
\tilde{B}_{k+1} = B_{k} - h\nabla \log\mathcal{L}_n (B_k) +\sqrt{2h}\,W_k,\qquad
k=0,1,\ldots.
\label{mala}
\end{align}
Note that $\tilde{B}_{k+1} $ is normally distributed with mean $ B_{k} - h\nabla \log\mathcal{L}_n (B_k)  $ and the covariance matrix equals to $ 2h\mathbf{I}_p $. This proposal is accepted or rejected according to the Metropolis-Hastings algorithm. That is the proposal is accepted with probabiliy:
$$
\min \left\lbrace 1, \frac{\mathcal{L}_n (\tilde{B}_{k+1}) q(B_k | \tilde{B}_{k+1}) }
{\mathcal{L}_n (B_k ) q(\tilde{B}_{k+1} | B_k ) } \right\rbrace,
$$
where 
$$
q(x' | x) \propto \exp \left(-\frac{1}{4h}\|x'-x +h\nabla \log\mathcal{L}_n (x) \|^2_F \right)
$$
is the transition probability density from $x$ to $x'$. The detail of the Metropolis-adjusted Langevin algorithm (denoted by MALA) for BRRR is given Algorithm \ref{mala_algoritm}. Compared to random-walk Metropolis–Hastings,  MALA has the advantage that it usually proposes moves into regions of higher probability, which are then more likely to be accepted.

\begin{remark}
Following \cite{roberts1998optimal}, the choice of the step-size $h$ is tuned such that the acceptance rate is approximate $0.5$. See Section \ref{sc_numerical} for some choices in special cases in our simulations.
\end{remark}

\begin{algorithm}{}
\caption{MALA for BRRR}
\begin{algorithmic}[1]
\State \textbf{Input}: matrices $Y\in\mathbb{R}^{n\times m} ,X \in \mathbb{R}^{n\times p} $
\State \textbf{Parameters}: Positive real numbers $\lambda,h,T$ 
\State \textbf{Onput}: The matrix $\hat{B} $
\State \textbf{Initialize}: $B_0 \gets (X^\top X + 0.1\mathbf{I}_p)^{-1}X^\top Y;  \hat{B} = \mathbf{0}_{p\times m}$ 
\For{$k = 1$ to $T$} 
    \State Simulate $\tilde{B}_{k}$ from \eqref{mala}
    \State Calculate $\alpha = \min \left\lbrace 1, \frac{\mathcal{L}_n (\tilde{B}_{k}) q(B_{k-1} | \tilde{B}_{k}) }
{\mathcal{L}_n (B_k ) q(\tilde{B}_{k} | B_{k-1} ) } \right\rbrace$
   \State Sample $u \sim U[0,1]$
   \If {$ u \leq \alpha $} 
   		\State  $B_k = \tilde{B}_{k}$ 
   \Else  
   \State  $B_k = B_{k-1}$ 
   \EndIf
    \State  $\hat{B} \gets  \hat{B} +  B_{k}/T $
\EndFor
\end{algorithmic}
\label{mala_algoritm}
\end{algorithm}

\section{Numerical studies}
\label{sc_numerical}
\subsection{Simulations setups and details}
First, we perform some numerical studies on simulated data to access the performance of our proposed algorithms. We consider the following model setups:
\begin{itemize}
\item Model I: A low-dimensional set up is studied with $n=100, p=12, m = 8$ and the true rank $r = {\rm rank}(B) = 3$. The design matrix $X$ is generated from $ \mathcal{N}(0, \Sigma) $ where the covariance matrix $\Sigma$ is with diagonal entries 1 and off-diagonal entries $\rho_X \geq 0 $. We consider $ \rho_X = 0 $ and $ \rho_X = 0.5 $, this creates a wide-range correlation in the predictors. The true coefficient matrix is generated as $B = B_1B_2^\top$ where $B_1 \in \mathbb{R}^{p\times r}, B_2 \in \mathbb{R}^{m\times r} $ and all entries in $ B_1 $ and $B_2$ are randomly sampled from $ \mathcal{N}(0, 1) $.

\item Model II: This model is similar to Model I, however, a high-dimensional set up is considered with $n=100, p=150, m = 90$.

\item Model III: An approximate low-rank set up is studied. This series of simulation is similar to the Model II, except that the true coefficient is no longer rank 3, but it can be well approximated by a rank 3 matrix:
$$
B = 2\cdot B_1B_2^\top + E,
$$
where $E $ is matrix with entries sampled from $ \mathcal{N}(0, 1) $.
\end{itemize}
Under each setting, the entire data generation process described above is replicated 100 times.

We compare our algorithms LMC, MALA to the naive reduced rank method (denoted RRR, see \cite{velu2013multivariate}) where the rank is selected by 10-fold cross validation. The RRR method is available from the R package 'rrpack'\footnote{\url{https://cran.r-project.org/package=rrpack}}. We also compare LMC and MALA to the Gibbs sampler from \cite{alquier2013bayesian}, however, we are just able to perform these comparisions in Model I as the Gibbs sampler is too slow for large dimensions, see Figure \ref{fg_runningtime}. The R codes for the Gibbs sampler are kindly provided by the author of the paper \cite{alquier2013bayesian}.

The evaluations are done by using the estimation error (Est) and the normalized mean square error (Nmse)
$$
{\rm Est} := \| B - \hat{B} \|^2_F/ (pm),
\quad
{\rm Nmse}:= \| B - \hat{B} \|^2_F/ \| B  \|^2_F ,
$$
that are calculated as the average of the mean squared errors from all 100 runs.  We also evaluate the average over 100 runs of the prediction error (Pred) as 
$$ 
{\rm Pred} := \| Y_{test} - X_{test}\hat{B} \|^2_F/ (nm)
$$ 
where $X_{test} $ is a newly generated $n\times p$ test-sample matrix of predictors and $Y_{test} $ is a newly generated $n\times m $ test-sample matrix of responses.  We also report the average of estimated rank,  denote Rank,  for different methods over all the runs.

The choice of the step-size parameters is set as: for Model III, we take $h = 3/(\sqrt{m}np)$; with Model II $h = 5/(mnp) $ and with Model I, $h = 2/(pm\sqrt{n})$. This choice is selected such that the acceptance rate of MALA is approximate 0.5. We fixed $\lambda = 3$ in all models. The LMC, MALA and Gibbs sampler are run with $T = 200$ iterations and we take the first $100$ steps as burn-in.

\subsection{Simulation results}

In low-dimensional setting where $pm<n$ as in Model I, Langevin MC algorithms (LMC, MALA) are able to recover the true rank of the model, see Table \ref{tb_model1}. The results of MALA are slightly better than LMC. The prediction errors of LMC and MALA are comparable to RRR and Gibbs sampler. In terms of other errors (Est and Nmse), LMC and MALA are twice worse than RRR and Gibbs sampler.

\begin{figure}[!ht]
\centering
\includegraphics[scale=.8]{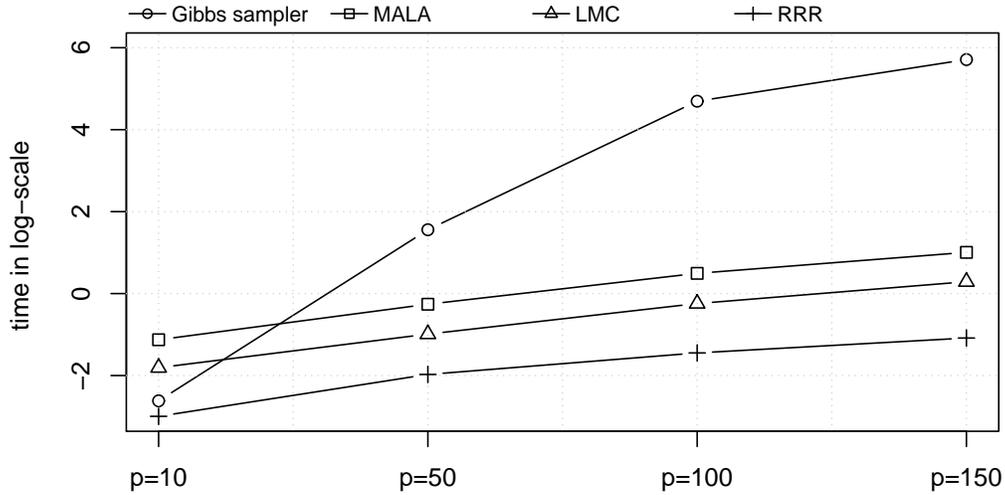}
\vspace*{-1cm}
\caption{Plot to compare the running times for 10 iterations of LMC, MALA Gibbs sampler and 10-fold cross validation RRR with fixed $n=100,m=90,r=2$ and the dimension $p$ is varied.}
\label{fg_runningtime}
\end{figure}

However, it is worth noting that the running time of our algorithms is linearly with $p$ where $n$ and $m$ are fixed, while the Gibbs sampler is not. More specifically, we conducted a comparision on the running time for these four algortihms where the dimension $p$ is varied by $10,50,100,150$ with fixed $n=100,m=90$. The results is given in Figure \ref{fg_runningtime}. It is clear that the Gibbs sampler is several magnitude slower than our algorithms. 

Results from high dimensional settings as in Model II and II reveal that our algorithms perform quite similar the RRR method, see Table \ref{tb_model23}, in term of all considered errors. Moreover, it is interesting that LMC and MALA are slightly better than RRR method in Model III where coefficient matrix is approximately low-rank. More specifically, MALA produces promissing results that are lightly better than LMC as well as RRR.

\begin{table}[ht]
\caption{Simulation results on simulated data in Model I for different methods, with their standard error in parentheses.  (Est: average of estimation error; Pred: average of prediction error; Rank: average of estimated rank).}
\centering
\small
\begin{tabular}{p{20mm}|cccc} 
\toprule\toprule
     & \multicolumn{4}{ c  }{$\rho_X = 0.0 $} 
 \\
    Errors    & LMC & MALA & RRR & Gibbs
 \\ \midrule
 $10^2\times$Est 	& 1.25 (0.21) & 1.26 (0.21) & 0.58 (0.14) & 0.59 (0.13)
               \\ 
Pred  			& 1.15 (0.06) & 1.15 (0.06) & 1.07 (0.05) & 1.07 (0.05)
			\\ 
$10^3\times$Nmse  & 4.66 (2.21) & 4.80 (2.46) & 2.16 (1.11) & 2.18 (1.06)
 \\
 Rank & 3 (0.0) & 3 (0.0) & 3 (0.0) & 3 (0.0)
 \\  
 \midrule
     & \multicolumn{4}{ c }{$\rho_X = 0.5 $} 
 \\  
 & LMC & MALA & RRR & Gibbs
 \\ \midrule
  $10^2\times$Est 	&  2.52 (0.41) & 2.39 (0.56) & 1.03 (0.24) & 1.12 (0.02)
  \\
 Pred  			& 1.17 (0.07) & 1.16 (0.07) & 1.08 (0.06) & 1.08 (0.07)
 \\
  $10^2\times$Nmse  & 0.94 (0.41) & 0.96 (0.55) & 0.43 (0.31) & 0.46 (0.29)
  \\
  Rank & 3 (0.0) & 3 (0.0) & 3 (0.0) & 3 (0.0)
 \\ \bottomrule\bottomrule
\end{tabular}
\label{tb_model1}
\end{table}

\begin{table}[ht]
\caption{Simulation results on simulated data in Model II \& III for different methods, with their standard error in parentheses.  (Est: average of estimation error; Pred: average of prediction error; Rank: average of estimated rank).}
\small
\centering
\begin{tabular}{p{25mm} | p{25mm}|ccc} 
\toprule\toprule
Models    & Errors    & LMC & MALA & RRR  
 \\ \midrule
           & Est 	& 1.00 (0.13) & 1.00 (0.13) & 0.99 (0.13) 
               \\ 
 II, $\rho_X = 0.0 $ & $10^{-2}\times$Pred  & 1.51 (0.23) & 1.51 (0.23) & 1.49 (0.23)
			\\ 
               & Nmse  & 0.34 (0.03) & 0.34 (0.03) & 0.33 (0.03)
               \\
   & Rank & 3 (0.0)        & 3 (0.0)      & 3 (0.0)   
\\ \midrule
           & Est 	& 1.01 (0.14) & 1.01 (0.14) & 0.99 (0.14) 
               \\ 
 II, $\rho_X = 0.5 $ &
  $10^{-1}\times$Pred  & 7.65 (1.20) & 7.65 (1.20) & 7.45 (1.20)
			\\ 
               & Nmse  & 0.34 (0.03) & 0.34 (0.03) & 0.33 (0.03)
               \\
   & Rank & 3 (0.0)        & 3 (0.0)      & 3 (0.0)   
\\ \midrule
               & Est 	& 4.32 (0.64) & 4.32 (0.64) & 4.34 (0.64) 
               \\ 
 III, $\rho_X = 0.0 $  &
 $10^{-2}\times$ Pred  & 6.53 (1.05) & 6.53 (1.05) & 6.56 (1.05)
			\\ 
               & Nmse  & 0.33 (0.03) & 0.33 (0.03) & 0.33 (0.03)
               \\
               & Rank & 3.04 (0.20) & 3.04 (0.20) & 3.04 (0.20)
\\ \midrule
               & Est 	& 4.41 (0.58) & 4.40 (0.57) & 4.43 (0.57) 
               \\ 
 III, $\rho_X = 0.5 $  &
 $10^{-2}\times$ Pred  & 3.31 (0.47) & 3.30 (0.47) & 3.32 (0.47)
			\\ 
               & Nmse  & 0.34 (0.03) & 0.34 (0.03) & 0.34 (0.03)
               \\
               & Rank & 3.09 (0.35) & 3.08 (0.34) & 3.08 (0.34) 
               \\ \bottomrule\bottomrule
\end{tabular}
\label{tb_model23}
\end{table}

\subsection{Real data application}
We apply our algorithms to a breast cancer dataset \cite{witten2009penalized} to acess its performance on real data set. This data consisting of gene expression measurements and comparative genomic hybridization measurements for $n = 89$ samples. The dataset is available from the R packge 'PMA' \cite{witten2009penalized}. This data were used before in \cite{chen2013reduced} in the context of reduced rank regression.

Following \cite{chen2013reduced}, we consider the gene expression profiles of a chromosome as predictors and the copy-number variations of the same chromosome as repsonse. The analysis is focused on chromosome $21 $, for which $m = 44$ and $p = 227 $. The data are randomly divided into a training set of size $n_{train} = 79$ and a test set of size $n_{test} = 10$. Model estimation is done by using the training data. Then the predictive performance is calculated on the test data by its mean squared prediction error $ \|Y_{test} - X_{test}\hat{B} \|_F^2 / (mn_{test}) $, where $ (Y_{test} , X_{test}) $ denotes the test set. We repeat the random training/test spliting process $100$ times and report the average mean squared prediction error and the average rank estimate for each method. The results are given in Table \ref{tb_realdata}, we can see that MALA is better than LMC.

\begin{table}[ht]
\caption{Comparision of the model fits to the real data. The mean suqared prediction errors (MSPE) and the estimated ranks are reported, with their standard error in parentheses.}
\centering
\begin{tabular}{p{17mm}|ccc} 
\toprule
     & LMC & MALA & RRR  
 \\ \midrule
MSPE	& 0.052 (.009) & 0.049 (.008) & 0.030 (.008) 
\\
Rank & 1.03 (.17) & 1.03 (.17) & 0.74 (.44)
            \\ \bottomrule
\end{tabular}
\label{tb_realdata}
\end{table}

\section{Discussion}
\label{sc_discuss}
It is noted that our Langevin MC approaches for BRRR are using a different prior on (approximate) low-rank matrix comparing with the matrix factorization as in Gibbs sampler. There are several other ways to define such priors on a whole matrix, see \cite{sundin2016bayesian}. For example, one could consider, with $\lambda >0$, 
$$ 
\pi (B) \propto \exp (\lambda {\rm Trace} (BB^\top)^{1/2} ) 
$$ 
where its log-prior is the nuclear norm that is also promoting the low-rank structure on $B$. The application of Langevin MC method for such priors would be interesting research directions in the future.

A vital part in the Langevin MC approach is choosing the step-size $h$. Here, in this work, the choice of $h$ is picked such that the acceptance rate in MALA is around $0.5$, motivating from \cite{roberts1998optimal}. We have tried with a decreasing step-size as $h_t = h_{t-1}/t$, however this choice does not improve the results at all compare to the choise defining through the acceptance rate. It is noted that there are several other way for choosing $h$, for example, $h$ is adaptively changed in each iteration as in \cite{marshall2012adaptive}. The study of such approach to BRRR  is left for future research.

Bayesian studies that incorporating sparsity into RRR model to account for both rank selection and variable selection have been recently carried out in \cite{goh2017bayesian,yang2020fully}. However, these works are still based on low-rank factorization priors and the implementation of the Gibbs sampler. Thus, the application of Langevin MC to this problem would be another interesting future work.

\section{Conclusion}
\label{sc_conclu}
In the paper, we have proposed efficient Langevin MC approach for BRRR. The performances of our algorithms are similar to the state-of-the-art method in simulations. More importantly, we showed that the proposed algorithms are significant faster than the Gibbs sampler. This is an interesting way that makes BRRR become more applicable in large data set.

\section*{Availability of data and materials}
The R codes and data used in the numerical experiments are available at:  \url{https://github.com/tienmt/BRRR} .

\section*{Acknowledgments}
This research of T.T.M was supported by the European Research Council grant no. 742158.

\bibliographystyle{apalike}


\end{document}